\documentstyle[preprint,aps,psfig]{revtex}
\title{ON THE CROSSOVER FROM BCS SUPERCONDUCTIVITY TO BOSE
CONDENSATION}
\author{  R. Micnas}
\address {Institute of Physics, 
A. Mickiewicz University,\\ ul. Umultowska 85, PL-61-614 Pozna\'n -- POLAND} 

\begin{document}
\maketitle 
\begin{abstract}
We outline a  microscopic approach to 
the superconducting fluctuations and pairing correlations in the attractive Hubbard model 
above $T_{c}$,  using  the functional integral method.
A crossover from BCS superconductivity to Bose condensation of preformed pairs 
is studied 
by constructing the  appropriate Ginzburg--Landau functionals. 
The differences between the  lattice and the continuum models are discussed. 
The case of quasi-two--dimensional superconductors as well as the model with
non-local pairing interaction are  also examined.
The effects of gaussian
 fluctuations of the order 
parameter are analyzed in the T-matrix approach,  the self--consistent 
Hartree approach and finally  within the Ginzburg--Landau theory. Simlarities
with a paramagnon theory of itinerant-electron magnetism are pointed out. 

\end{abstract}
\draft
\newpage 
\section{Introduction.}

The problem of a crossover from a weak coupling BCS superconductivity to Bose
condensation of preformed local pairs has recently  attracted  great
attention, stimulated by  the 
experimental results  cocerning unusual properties of new
high $T_c$ superconductors as well as  the discovery of Bose
condensates in optically trapped dilute atomic gases.
It has been established that high temperature superconductors (HTS)
generally  exhibit
a low carrier density,  small value of the Fermi energy 
$(\propto 0.1-0.3$~eV), short coherence length $\xi_{0}$ and they are 
{\em extreme type II superconductors} 
with the Ginzburg-Landau parameter
$\kappa \gg 1$\cite{micnas90,uemura91,schneider92,NATO}. Experimental studies of HTS and also other
non conventional superconductors 
indicate that the cuprate high $T_c$, bismuthates, fullerenes
and Chevrel phases belong to a unique group of superconductors
characterized by high transition temperatures relative to the values
of $n_s/m^{*}$. These materials have their $T_c$ proportional
to $T_F$ (the Fermi temperature) or $T_B$ (the Bose-Einstein
condensation temperature)\cite{uemura91}, with $T_B\approx (3-30)T_c$ and
$T_F\approx (10-100)T_c$. Moreover, the existence of the pseudogap phase in the underdoped HTS is well experimentally
established and it can be interpreted as 
a state with pairing without long-range phase coherence \cite{NATO,orenstein}.
The above points show that these materials are not in a weak
coupling BCS limit, but rather in the intermediate to extreme strong
coupling regime. 

One of the central questions regarding the physics
of HTS is the understanding of the evolution from a weak coupling limit of
large Cooper pairs to the strong coupling regime of small local pairs
with increasing coupling constant and description of the  intermediate
(crossover) regime.

Several attempts to describe the crossover have employed the BCS mean field
theory, generalized to allow a change in the chemical potential due to
the pairing away from the weak coupling limit
\cite{eagles,leggett,rob81,noz85,miyake,ran}.
It has been shown that the BCS wave function for $T=0$K is a reasonable ansatz
for any attraction strength. For the fermions in  continuum, the  controlling parameter
is $\xi_{0}k_{F}$ ($\xi_{0}$ - the coherence length , $k_{F}$ - the Fermi
momentum) and the BCS and BC limits correspond to $~\xi_{0}k_{F}\gg
1$ and $ \xi_{0}k_{F}\ll 1$, respectively.  An extension to finite
temperatures has been first proposed by Nozier\'{e}s and Schmitt-Rink
\cite{noz85}, who  included single-pair fluctuations within a
T-matrix approximation and  found   that  $T_{c}$ for the 3D continuum
dilute fermion system evolves smoothly between the limits (see also Refs.
\cite{melo,Haussmann,chicago}). Similar analysis has
also been carried out for a 2D dilute attractive Fermi gas   
 \cite{Varma et al,Traven}.

In this paper we overview the functional integral approach to the crossover
problem using mainly the Hubbard model with on-site attraction.
The attractive Hubbard model is a generic   model
of superconductivity on a lattice which can describe the evolution from
the BCS type superconductivity to a Bose condensation of 
real space pairs, with increasing coupling strength 
(for review see Refs.\cite{micnas90,randeria94,TKRM}). 
The layout of this paper is the following.
In Section II,  we present application of the  
functional integration  method to  the negative U Hubbard model. 
In Section III, the crossover is studied by 
a detailed derivation of 
the Ginzburg--Landau theory in  the weak attraction limit 
and the corresponding Ginzburg--Pitaevskii functional  in  
the strong coupling regime.
We discuss  the differences 
between the lattice and continuum fermion models with s-wave attraction. 
The case of quasi-2D-superconductivity and the model
with pair hopping interaction are also discussed. 
In Section IV,  we discuss the Gaussian fluctuations and the T-matrix approach.
The self-consistent approach to the superconducting fluctuations and pairing
correlations 
is further discussed in Section V, where 
a derivation of  a self--consistent Gaussian theory is given.
In Section VI we analyze the amplitude fluctuations 
above $T_{c}$ using the Ginzburg--Landau theory. 

\section{Functional Integral Approach.}   
Here we will outline the functional integral approach to 
the attractive Hubbard model. 
The  Hamiltonian of the model can be split into two parts
\begin{eqnarray}\label{Hubbard}
\label{Hamiltonian}
H=H_{0}+H_{1}, \\
H_{0}=\sum_{i,j,\sigma}\hat{t}_{ij}c_{i\sigma}^{\dagger}c_{j\sigma}\\
H_{1}= -|U|\sum_{i}\rho_{i}^{\dagger}\rho_{i},
\end{eqnarray} 
\noindent 
where $\rho_{i}^{\dagger}=c^{\dagger}_{i\uparrow}c^{\dagger}_{i\downarrow}$,
and $\hat{t}_{ij}=t_{ij}-\mu \delta_{ij}$. $t_{ij}$ is the transfer integral,
$|U|$-the on-site attraction and $\mu$ stands for the chemical potential. 
The grand partition function of the model (\ref{Hubbard}) in the interaction representation 
with respect to ${{H_{0}}}$ is given by 
\begin{eqnarray}\label{z}
Z  =   Tr \exp[-\beta (H_{0}+H_{1})]=
Tr[ e^{-\beta H_{0}}T_{\tau}\exp(-\int_{0}^{\beta}d\tau H_{1}(\tau))] = 
\nonumber\\
Z_{0}\left\langle T_{\tau}\exp(-\int_{0}^{\beta}d\tau H_{1}(\tau))\right\rangle_{0}~~,
\end{eqnarray} 
\noindent 
where $T_{\tau}$ is the "time" ordering operator, 
$H_{1}(\tau)=\exp(\tau  H_{0})H_{1}\exp(-\tau H_{0})$,$ Z_{0} = 
Tre^{-\beta H_{0}},~~\langle\cdots\rangle_{0} = 
Tr(\cdots e^{-\beta H_{0}})/(Tre^{-\beta H_{0}})$ and $\beta =(k_{B}T)^{-1}$.
The functional integral representation of $Z$ in terms 
of auxiliary fields ${\Delta_{i}}(\tau)$ is obtained in a 
standard manner by making use of the 
Hubbard--Stratonovich transformation 
\cite{Hubbard} 
\begin{equation}
\label{part1}Z = Z_{0}\int D^{2}{\Delta}exp(-\Psi[\Delta_{i}(\tau)]),
\end{equation} 
\noindent
where
\begin{eqnarray}
\label{part2}
\Psi[\Delta_{i}(\tau)] = 
\sum_{i}\int_{0}^{\beta} d\tau|\Delta_{i}(\tau)|^{2}/|U| ~ -  ~ \Omega[\Delta],
\\
\Omega[\Delta]=-\ln\left\langle T_{\tau}
\left[exp(\sum_{i}\int_{0}^{\beta}d\tau
[\Delta_{i}(\tau)\rho_{i}^\dagger(\tau)+H.c])\right]\right\rangle_{0}~~,
\end{eqnarray} 
$ D^{2}{\Delta}=D{\Delta}D{\Delta}^{*}.$
Introducing the Nambu representation 
$\Psi^{\dagger}_{i}=\left(c^{\dagger}_{i\uparrow},c_{i\downarrow}\right)$,
one obtains for $\Omega[\Delta]$
\begin{eqnarray}
\label{part3}
\Omega[\Delta]=- \ln\left\langle T_{\tau}
\left[exp(\sum_{i}\int_{0}^{\beta}d\tau\Psi^{\dagger}_i\hat\Delta_i(\tau)
\Psi_i(\tau))\right]\right\rangle_{0}~~,
\end{eqnarray}
\noindent
where $\hat{\Delta_{i}}=\left(\begin{array}{cc}0 & \Delta_{i}\\\Delta_{i}^{*} & 0 \end{array}\right).$  
The matrix Green's function 
\begin{equation}
\hat{\cal{G}}_{i,j}(\tau,\tau')= 
-\langle T_{\tau}\Psi_{i}(\tau)\Psi^{\dagger}_{j}(\tau')\rangle =
\left(\begin{array}{cc}G_{ij}(\tau,\tau') & F_{ij}(\tau,\tau')\\
F_{ij}^{\dagger}(\tau,\tau') & -G_{ij}(\tau,\tau')\end{array}\right)~~,
\end{equation} 
\noindent
where the field operators are given 
in the Heisenberg representation with respect to the Hamiltonian
(\ref{Hamiltonian}),
can be evaluated within the functional integral scheme as follows 
\begin{equation}
\hat{\cal {G}}_{i,j}(\tau,\tau')=\left\langle\hat{\cal {G}}_{i,j}(\tau,\tau',
\{\Delta\})\right\rangle_{\Delta}~~,
\end{equation} 
\noindent
where the average 
$\left\langle \cdots \right\rangle_{\Delta}=
\int D ^{2}\Delta \exp(-\Psi[\Delta])
(\cdots)/\int D^{2}\Delta \exp(-\Psi[\Delta])$       
denotes the functional average. 
The  Green's function: $\hat{\cal{G}}_{i,j}(\tau,\tau',\{\Delta\})$ is just 
the Green function for electrons in the random fields $\Delta_{i}(\tau)$ and is 
given 
as 
\begin{eqnarray}\label{Green}
\hat{\cal{G}}_{i,j}(\tau,\tau',\{\Delta\})=
-\frac{\left\langle T_{\tau}\Psi_{i}(\tau)\Psi^{\dagger}_{j}(\tau')S(\lambda)\right\rangle_{0}}
{\left\langle S(\lambda)\right\rangle _{0}}\nonumber\\
=
-\left\langle T_{\tau}\Psi_{i}(\tau)\Psi^{\dagger}_{j}(\tau')
\right\rangle_{H_{0}+H_{r}}~~.\\S(\lambda)=
T_{\tau}exp\left(-\lambda\int_{0}^{\beta} d\tau\Psi^{\dagger}
\hat{\Delta}_{i}(\tau)\Psi_{i}(\tau)\right)~~,
\end{eqnarray} 
\noindent
$\left\langle\cdots\right\rangle_{H_{0}+H_{r}}$ denotes the average with the Hamiltonian $H_{0}+H_{r}$ 
with $H_{0}$ given by Eq.(1) and $H_{r}$ given below
\begin{equation}
H_{r}=-\lambda\sum_{i}\left(\rho_{i}^{\dagger}\Delta_{i}(\tau)+H.c.\right)~~,
\end{equation} 
\noindent
and $\lambda$ is a 
formal parameter being set to 1 at the end.
This local Green's function satisfies the BCS equation in the presence 
of spatially and 
time varying local order parameter: 
\begin{eqnarray}
\label{local}\sum_{l}\left(-\frac{\partial}{\partial\tau}\delta_{il}-\hat{t}_{il}
\sigma_{z}-\lambda\hat{\Delta}_{i}(\tau)\right)\hat{{\cal G}}_{lj}(\tau,\tau',\{\Delta\}) =
\hat{1}\delta_{ij}\delta(\tau-\tau')~~,
\end{eqnarray}
\noindent
where $\sigma_{z}$ is the Pauli matrix. 
An exact representation for the partition function can be given by 
\begin{eqnarray}
\label{exact}
Z=Z_{0}\int D^{2}\Delta exp[-\Psi]\nonumber\\
\Psi=\sum_{i}\int_{0}^{\beta}d\tau|\Delta_{i}(\tau)|^{2}/|U| 
-Tr\ln(1-\hat{\Delta}\hat{\cal G}_{0}),
\end{eqnarray} 
\noindent
where
\begin{equation}\hat{\cal G}^{-1}_{0}=\left(-\frac{\partial}{\partial\tau}-
\hat{t}_{ij}\sigma_{z}\right)\delta_{ij}\delta(\tau-\tau')
\end{equation}
 We note that formal power series expansion of $\Psi$ will contain only
 even powers in $\Delta$ and it has the form 
\begin{equation}
\label{exp1}Tr\ln(1-\hat{\Delta}\hat{\cal G}_{0})=
-\sum_{k=1}^{\infty}\frac{Tr\left(\hat{\Delta}\hat{\cal G}_{0}\right)^{2k}}{2k}~~.
\end{equation}
The explicit form of the expansion up to the 4th order 
is given below 
\begin{eqnarray}
\label{exp}
\Psi^{(4)}=
\sum_{{\bf q},n}\left(\frac{1}{|U|}+\chi({\bf q},i\varepsilon_{n})\right)|
\Delta_{\bf q}(i\varepsilon_{n})|^{2} \nonumber \\+
\frac{1}{2N\beta}\sum_{{\bf p_{1}p_{2}p_{3}},n_{1}n_{2}n_{3}}\Gamma^{+-+-}({\bf p_{1},p_{2},p_{3}};n_{1},n_{2},n_{3})
\Delta^{*}_{\bf p_{1}}(i\varepsilon_{n_{1}})\Delta^{*}_{\bf p_{2}}(i\varepsilon_{n_{2}})\nonumber \\
\Delta_{\bf p_{3}}(i\varepsilon_{n_{3}})\Delta_{\bf p_{1}+p_{2}-p_{3}}(i\varepsilon_{n_{1}}+
i\varepsilon_{n_{2}}-i\varepsilon_{n_{3}})~~,
\end{eqnarray} 
In the above equation 
\begin{eqnarray}
\label{free}
\chi({\bf q},i\varepsilon_{n})=
\frac{1}{N}\sum_{k}\frac{f(\varepsilon_{k}-
\mu)-f(-\varepsilon_{k+q}+\mu)}
{\varepsilon_{k}+\varepsilon_{k+q}-2\mu -i\varepsilon_{n}}~~,
\end{eqnarray} 
\noindent 
is the free particle pairing susceptibility, 
$f(x)=\left(\exp(\beta x)+1\right)^{-1}$ is the Fermi 
distribution function,
$\varepsilon_{n}=\frac{2\pi}{\beta}n$  
are the bosonic Matsubara 
frequencies. $\Delta_{\bf k}(i\varepsilon_{n})$ is 
the Fourier transform of the field 
$\Delta_{i}(\tau)$ and is given by 
\begin{eqnarray}
\Delta_{{\bf q}}(i\varepsilon_{n})=
\frac{1}{\sqrt{\beta N}}\sum_{j}
\int_{0}^{\beta}d\tau 
e^{i\left(\varepsilon_{n}\tau-{\bf q}\cdot{\bf R}_{j}\right)}\Delta_{j}(\tau).
\end{eqnarray} 
The quartic term is of the form
\begin{eqnarray}
\Gamma^{+-+-}({\bf p_{1},p_{2},p_{3}};n_{1},n_{2},n_{3})=
\frac{1}{\beta N}\sum_{{\bf k},m}{\cal G}^{+}_{0}(k,i\omega_{m}){\cal G}^{-}_{0}(k+p_{1},i\omega_{m}-
i\varepsilon_{n})\nonumber\\
{\cal G}^{+}_{0}(k+p_{3}-p_{2},i\omega_{m}+i\varepsilon_{n_{2}}-i\varepsilon_{n_{3}})
{\cal G}^{-}_{0}(k+p_{3},i\omega_{m}-i\varepsilon_{n_{3}}),
\end{eqnarray} 
\noindent
where 
$\hat{\cal G}_{0}^{-1}=i\omega_{n}-(\varepsilon_{k}-\mu)\sigma_{z}$
is the Fourier transform of the free fermion Green's function and 
$\omega_{n}=\frac{2\pi}{\beta}(n+\frac{1}{2})$ are the fermionic 
Matsubara frequencies. 
If  we neglect all the frequency dependence of 
$\Delta$ in the expansion(\ref{exp1}) (the static approximation), 
then we have the following result  for the quartic term 
(we also neglect the momentum dependence):
 \begin{equation}
 \Gamma^{+-+-}(0,0,0)=\frac{b}{2}=
 \frac{1}{\beta N}\sum_{{\bf k} m}\left[\frac{1}{(i\omega_{m})^2-
 \bar{\varepsilon_{k}}^2}\right]^2,
 ~~~\bar{\varepsilon_{k}}=\varepsilon_{k}-\mu~~,
 \end{equation}
  \noindent 
  which is the standard expression for the quartic term in the Ginzburg-Landau (G-L)
  theory.  
  We note that the corresponding coefficients can be evaluated explicitly and 
  basing on such expansion one can study the question of the crossover from the 
  BCS pairing 
  (when $\mu\approx E_{F}$) to the bosonic limit of tightly bound pairs 
  $(\mu\approx-E_{b}/2$, where $E_{b}$ is the binding energy for 
  two fermions in an empty lattice).
   For  completeness, we give below the time-independent Ginzburg-Landau functional
 \begin{equation}
 \bar{\Psi}/\beta=\sum_{{\bf q}}\left({\bar a}+c\frac{q^2}{2m}\right)|\Delta_{q}|^2+\frac{b}{2}
 \sum_{k_{1},k_{2},k_{3}}\Delta^{*}_{k_{1}}\Delta^{*}_{k_{2}}\Delta_{k_{3}}\Delta_{{k_{1}+k_{2}-k_{3}}}~~,
 \end{equation} 
 \noindent 
 where $\Delta_{q}=\Delta_{{\bf q}}(0)$. 
 The coefficients appropriate to a weak coupling superconductor are given by 
 \begin{eqnarray}
 \label{GLcoefficients}
 {\bar a}=N(E_{F})\ln\frac{T}{T_{c}},~
 c=\frac{7\zeta(3)N(E_{F})E_{F}}{12\pi^2(k_{B}T_{c})^{2}},~ 
 b=\frac{7\zeta(3)N(E_{F})}{8\pi^2(k_{B}T_{c})^{2}}~~,
 \end{eqnarray} 
 where $N(E_{F})$ is the density of states (DOS) at the Fermi level and $E_{F}$
 is the Fermi energy.
 The above G-L functional will be used in Section VI in discussing 
 the effects of amplitude fluctuations above $T_{c}$. 
\section{From BCS to Bose superconductivity}
As mentioned above, given the functional integral representation 
(\ref{exact}), the evolution from BCS to Bose 
type superconductivity can be analyzed.
The  coefficient in quadratic term of (\ref{exp}) is of the form
\begin{eqnarray}
A({\bf q},i\varepsilon_{n})=
\frac{1}{|U|}+\chi({\bf q},i\varepsilon_{n}).
\end{eqnarray}
It is convenient to eliminate the pairing potential $|U|$ in favor of the
two-particle binding energy in vacuum, $E_{b}$,\cite{leggett}, through the
relation
\begin{eqnarray}
\label{boundstate}
\frac{1}{|U|}=\frac{1}{N}\sum_{k}\frac{1}{2\varepsilon_{k} +E_{b}}.
\end{eqnarray}
In the long-wave limit, 
assuming the parabolic spectrum for fermions with the mass $m$,  
the expansion is given by 
\begin{eqnarray}
A({\bf q},i\varepsilon_{n})=
{\bar a} +c\frac{{\bf q}^{2}}{2m} -i{\bar d}\varepsilon_{n},
\end{eqnarray}
where
\begin{eqnarray}\label{coeficients}
{\bar a}=\frac{1}{N}\sum_{k}\frac{1}{2\varepsilon_{k} +E_{b}}-
\frac{1}{N}\sum_{k}\frac{\tanh(\beta{\bar \varepsilon}_{k}/2)}
{2{\bar \varepsilon}_{k}},\\
c=\frac{1}{N}\sum_{k}\{\left(\frac{\tanh(\beta{\bar \varepsilon}_{k}/2)}
{4{\bar \varepsilon}^2_{k}}-\frac{\beta}
{8{\bar \varepsilon}_{k}\cosh^2(\beta{\bar \varepsilon}_{k}/2)}\right)\nonumber\\
+\left(\frac{\beta^2\tanh(\beta{\bar \varepsilon}_{k}/2)}
{\cosh^2(\beta{\bar \varepsilon}_{k}/2)}
+\frac{\beta}{{\bar \varepsilon}_{k}\cosh^2(\beta{\bar \varepsilon}_{k}/2)}
-\frac{2\tanh(\beta{\bar \varepsilon}_{k}/2)}{{\bar \varepsilon}^2_{k}}\right)
\frac{({\bf k}\cdot{\bf n})^2}{8m {\bar\varepsilon}_{k}}\},\\
\label{coeffb}
b/2=\frac{1}{N}\sum_{k}\left( \frac{\tanh(\beta{\bar \varepsilon}_{k}/2)}
 {4{\bar \varepsilon}^3_{k}}-\frac{\beta}{8{\bar \varepsilon}^2_{k}}
 \frac{1}{\cosh^2(\beta{\bar \varepsilon}_{k}/2)}\right),
\end{eqnarray}
and the coefficient ${\bar d}$ will be specified later.

In the BCS/Fermi liquid regime $E_{b}/E_{F}\ll 1$, $\mu \approx E_{F} \gg T$ and 
we  recover the standard form of
the Ginzburg Landau functional:
\begin{eqnarray}
S_{GL}=\sum_{q}\left({\bar a}+c\frac{q^2}{2m}-i{\bar d}\varepsilon_{n}\right)
|\Delta_{q}|^2 +\frac{b}{2}\sum_{1,2,3}\Delta^*_{1}\Delta^*_{2}
\Delta_{3}\Delta_{1+2-3},
\end{eqnarray}
$q=({\bf q},i\varepsilon_{n})$, with the coefficients ${\bar a},b,c$  given by Eqs.(\ref{GLcoefficients}) 
and with ${\bar d}=\pi N(E_{F})/8k_{B}T_{c}$. Indeed, with the rescaling 
$\Phi(k)=\sqrt{2c}\Delta_{k}$ (the Gorkov relation),
 it takes the conventional form of the time-dependent G-L functional.  

In the opposite Bose limit $E_{b}/E_{F}\gg 1$, and the chemical potential$ -\mu $
approaches $E_{b}/2$ at the transition point, thus $-\mu \gg T_{c}$. 
Defining the effective  chemical potential $\mu ={\bar \mu} -\frac{E_{b}}{2}$, 
we then obtain, 
${\bar a}=-\frac{N(0)}{2E_{b}}2{\bar \mu}, 
c=\frac{N(0)}{4E_{b}}, {\bar d}=\frac{N(0)}{2E_{b}}$, 
where  $N(0)$ is  the DOS in 2D.
After the rescaling of the pair-field amplitude
$\Phi({\bf q},i\varepsilon_{n})=\sqrt{\frac{N(0)}{2E_{b}}}
\Delta_{{\bf q}}(i\varepsilon_{n})$,
the effective action becomes

\begin{eqnarray}\label{Gross}
S_{eff}=\sum_{q}\left(\frac{q^2}{2m^{*}}-2{\bar \mu}-i\varepsilon_{n}\right)
|\Phi(q)|^2 +{\bar b}\sum_{1,2,3}\Phi^*(1)\Phi^*(2)\Phi(3)\Phi(1+2-3),
\end{eqnarray}
where  $m^{*}=2m, {\bar \mu}=E_{b}/2+\mu$, 
and ${\bar b}$ is the parameter describing
the interaction between the  tightly bound pairs.
Eq.(\ref{Gross}) yields  required  Gross-Pitaevski functional describing
interacting bosons in the dilute limit, 
with the effective mass $m^{*}=2m$ and the effective chemical potential 
${\bar \mu}$\cite{zwerger,pieri}.  
The parameter ${\bar b}$ describes repulsive interaction between the bosons and 
it strongly depends on dimensionality.
For a  constant DOS, the above procedure 
yields 
\begin{eqnarray}
\label{bcoeff}
b/2=\frac{N(0)}{16}\frac{1}{|\mu|^2},
\end{eqnarray}
where  $N(0)=\frac{m}{2\pi}$.
In the limit $|\mu|\rightarrow E_{b}/2$, the rescaling gives 
${\bar b}\approx 2/N(0)$. Thus, the only interaction is due to the Pauli
principle for constituent fermions.  However, in general, 
one would expect, that the effective boson-boson repulsion should also be
controlled by  
the scattering length for two bosons.
In  the strictly d=2 case, the scattering amplitude at long-waves and 
low-frequencies vanishes,
so the above mapping must be taken with care, and the effects of higher orders
in the expansion (\ref{exp}) may be of importance. 
We also notice that in 2D, the coefficient $b$ of the GL functional 
(Eq.(\ref{bcoeff})),
 when $\mu \rightarrow 0$, is divergent, thus indicating  a 
singular point in the evolution from
BCS to Bose limit. This however can be again cured by taking into account the
interaction between the fluctuations\cite{Yamada}. 

A question of the derivation of the time-dependent Ginzburg-Landau 
functional by this method,  in the weak and strong coupling
regimes, is discussed in Ref.\cite{melo}.

Finally, we should mention that the effective mass of the bosons $m^{*}$ in the
continuum limit approaches 2m. If 
 the tight binding spectrum of the form
 $\epsilon_{k}=-2t(cos(k_{x}a)+cos(k_{y}a)+cos(k_{z}a))$, $a$ being the lattice constant, is used, 
 one obtains the following expansion for $A({\bf q},i\varepsilon_{n})$:
 \begin{eqnarray}
 A({\bf q},i\varepsilon_{n})={\bar a} +C{\bf q}^2 -id\varepsilon_{n},
  \end{eqnarray}
  where ${\bar a}$ and $b$ are  given by Eq.(28) and Eq.(31) and 
  \begin{eqnarray}\label{expnegU} 
C=\frac{1}{2N}\sum_{k}\left(\frac{f'(\bar{\varepsilon}_{k})}{2\bar{\varepsilon}_{k}}
+\frac{1-2f(\bar{\varepsilon}_{k})}{4\bar{\varepsilon}_{k}^2} \right) 2ta cos(k_{x}a)\nonumber\\
+\frac{1}{2N}\sum_{k}\left(\frac{f''(\bar{\varepsilon}_{k})}{2\bar{\varepsilon}_{k}}
-\frac{2f'(\bar{\varepsilon}_{k})}{4\bar{\varepsilon}_{k}^2}
-\frac{1-2f(\bar{\varepsilon}_{k})}{4\bar{\varepsilon}_{k}^3}\right)
4t^2a^2sin^2(k_{x}a),
\end{eqnarray}
$f^{n}$ are the derivatives of the Fermi function.
The coefficient ${\bar d} $ can be  obtained by first making an analytical continuation
 $ \chi({\bf {q}}=0,i\varepsilon_{n})\rightarrow
 \chi({\bf {q}}=0,\omega +i0^{+}) 
 $ and then expanding in $\omega$\cite{melo}.
  One gets:
 \begin{eqnarray}\label{coeffd}
 {\bar d}=\frac{1}{N}\sum_{k}\frac{1-2f(\bar{\varepsilon}_{k})}{4\bar{\varepsilon}_{k}^2}
 +i\frac{\pi N(\mu)\beta}{8}\Theta(\mu),
 \end{eqnarray}
 where $\Theta(\mu)$ is the Heaviside step function.
In the Bose limit the coefficients take the form:
${\bar a} = -2\bar{\mu}/{U^2}, C = 2t^2a^2/|U|^3, d = 1/U^2$, and upon
rescaling the pair-field amplitudes one gets the functional (32), with    
the bosonic mass $m^{*}=|U|/(4t^2a^2)$. This result is in accord with the
perturbation expansion for the attractive Hubbard model in the strong coupling
limit\cite{micnas90}.

\subsection{Quasi-two dimensional case}
 Let us now consider the quasi-two dimensional case and assume the following 
 spectrum for the fermions interacting via the on-site attractive potential $U$:
\begin{eqnarray}
 \varepsilon_{k}=\frac{k^2}{2m_{\parallel}}-2t_{\perp}\cos(k_{z}d),
 \end{eqnarray}
where $k^2=k_{x}^2+k_{y}^2$, $m_{\parallel}$ is the in - plane fermion mass and $t_{\perp}$ is
the hopping amplitude between the neighboring planes at distance $d$.
In contrast to the strictly 2D case, now the $s$-wave BCS  pairing 
and the bound state
formation are not necessarily  related to each other.
The bound state condition Eq.(\ref{boundstate}), shows that for 
a given ratio of $t_{\parallel}/t_{\perp}$ the threshold 
 must be reached for $|U|$, 
  to form the bound state in 
 vacuum.
 In the BCS case, evaluation of 
 the quadratic and quartic coefficients will yield an appropriate
 generalization of the G-L functional to the  quasi 2D case.
 
 Let us consider the strong coupling case above the 
 threshold to form the bound pair,
 and where $\mu$ is below the
 bottom of the fermionic band and the temperature region $k_{B}T<E_{b}$.
In such a case we can neglect the Fermi factors in the evaluation of the
quadratic coefficient and obtain
\begin{eqnarray}
A({\bf q},i\varepsilon_{n})= 
\frac{1}{N}\sum_{k}
\frac{1}{\frac{k^2}{m_{\parallel}}-4t_{\perp}\cos(k_{z}d)+
4t_{\perp}+E_{b}}-\nonumber\\
\frac{1}{N}\sum_{k}
\frac{1}{\frac{k^2}{m_{\parallel}}+
\frac{q^2}{4m_{\parallel}}-4t_{\perp}\cos(q_{z}d/2)
\cos(k_{z}d)+4t_{\perp}-2\mu-i\varepsilon_{n} },
\end{eqnarray} 
and ${\bf q}=(q,q_{z})$.
   
After integrating over $k$, one gets:

\begin{eqnarray}
\label{q2d}
A({\bf q},i\varepsilon_{n})=
\frac{N(0)}{2}\ln\left(\frac{\frac{q^2}{4m_{\parallel}}-2\mu-i\varepsilon_{n} +
\sqrt{\left(\frac{q^2}{4m_{\parallel}}-2\mu-i\varepsilon_{n}\right)^{2}
-16t^{2}_{\perp}\cos^{2}(\frac{q_{z}d}{2})}}
{E_{b}+4t_{\perp} +\sqrt{\left(E_{b}+
4t_{\perp}\right)^{2}-16t^{2}_{\perp}}}\right).    
\end{eqnarray} 

Introducing the  effective chemical potential by the relation
$\mu=-\frac{1}{2}E_{b}-2t_{\perp}+{\bar \mu}$, and expanding (\ref{q2d}) 
to lowest order in $q_{\parallel}$ and $i\varepsilon_{n}$, we obtain

\begin{eqnarray}
A({\bf q},i\varepsilon_{n})=
\frac{N(0)}{2E_{b}}\left(\frac{q^2}{4m_{\parallel}} +\frac{(4t_{\perp})^2}
{8E_{b}}(1-\cos(q_{z}d)) -2{\bar \mu} -i \varepsilon_{n}\right).
\end{eqnarray}

By rescaling the pairing-field amplitude $\Phi({\bf q},i\varepsilon_{n})=
\sqrt{\frac{N(0)}{2E_{b}}}\Delta_{{\bf q}}(i\varepsilon_{n})$, 
one obtains the Gross-Pitaevski 
action for the quasi-2D-bosons 
\begin{eqnarray}
S_{eff}=\sum_{q}\left(\frac{q^2}{4m_{\parallel}}+\frac{1}{2m^2_{\perp}d^4E_{b}}
(1-\cos(q_{z}d))-
2{\bar \mu}-i\varepsilon_{n}\right)
|\Phi(q)|^2 +\nonumber\\
{\tilde b}\sum_{1,2,3}\Phi^*(1)\Phi^*(2)\Phi(3)\Phi(1+2-3),
\end{eqnarray}
where $m_{\perp}=1/t_{\perp}^2d^2$ is the fermionic mass in the perpendicular
direction.
The bosons have the in-plane mass $2m_{\parallel}$ and the perpendicular  
mass 
$m^{*}_{\perp}=2E_{b}m^2_{\perp}d^2$ being much greater than $m_{\perp}$.
The physical origin of the increased perpendicular mass is clear, to move the
pair between the planes one has to break it first. 
Indeed, these (composite) bosons will condense at the temperature given by\cite{micnas92}
\begin{eqnarray}
k_{B}T_{c}=\frac{\pi n^{*}d }{2
m_{\parallel}\ln(2k_{b}T_{c}\nu E_{b}m^2_{\parallel}d^4)}\approx
\frac{E_{F}}{2\ln((4k_{b}T_{c}E_{b}m^2_{\parallel}d^4)},
\end{eqnarray}
$n^{*}=n/2$ is the boson density, $n$ being the  electron density
and $E_{F}=\pi n d/m_{\parallel}$.
Thus, the transition temperature for  the quasi-two dimensional case in 
 the preformed pair regime is proportional 
to $E_{F}$. This temperature will  decrease with increasing interaction making 
the pair binding stronger.
We should point out that independently, same result has been obtained in 
Ref.(\cite{ukrainians}), by using a different approach.

\subsection{The extended Hubbard model with pair-hopping interaction}
The above derivations of the Ginzburg-Landau functionals can be extended  to
the case of the Hubbard model with the pair hopping interaction (the 
Penson-Kolb-Hubbard
model), which involves 
the non-local pairing interaction and  is given
by the Hamiltonian:
\begin{eqnarray}\label{pairhopping}
{\tilde H}=\sum_{i,j,\sigma}\hat{t}_{ij}c_{i\sigma}^{\dagger}c_{j\sigma}
+U\sum_{i}\rho_{i}^{\dagger}\rho_{i}-
\frac{1}{2}\sum_{i,j}(J_{ij}\rho^{\dagger}_{i}\rho_{j}+\rm {H.c.})~,
\end{eqnarray}
where $J_{ij}$ is the pair-hopping (intersite charge exchange)
parameter\cite{rob}.
The functional integral representation of the partition function of the model
(\ref{pairhopping}) is given by  Eq.(\ref{exact}) with
\begin{eqnarray}\label{pairhopfunc}
\Psi=\sum_{{\bf q},n}\frac{|
\Delta_{q}(i\varepsilon_{n})|^{2}}{V_{\bf q}}
-Tr\ln(1-\hat{\Delta}\hat{\cal G}_{0}),
\end{eqnarray}
where $V_{\bf {q}}=-U + J_{\bf {q}}$ and $J_{\bf {q}}=
2J(\cos(k_{x}a)+\cos(k_{y}a)+\cos(k_{z}a))$ is the Fourier transform of
$J_{ij}$.
The coefficient of the quadratic term in the power series expansion of $\Psi$ in
$\Delta$ is 
\begin{eqnarray}
A({\bf q},i\varepsilon_{n})=
\frac{1}{V_{\bf {q}}}+\chi({\bf q},i\varepsilon_{n}),
\end{eqnarray}
where the pairing susceptibility is determined from Eq.(\ref{free}).
The small ${\bf {q}},\omega $ expansion is now of the following form 
\begin{eqnarray}
 A({\bf q},i\varepsilon_{n})={\bar a_{1}} +C_{1}{\bf q}^2 -id\varepsilon_{n},
\end{eqnarray}
  where 
  \begin{eqnarray}
  {\bar a}_{1}=\frac{1}{V_{0}}-\frac{1}{N}\sum_{k}\frac{\tanh(\beta{\bar \varepsilon}_{k}/2)}
{2{\bar \varepsilon}_{k}}=
  \frac{1}{N}\sum_{k}\frac{1}{2\varepsilon_{k} +E_{b}}-
\frac{1}{N}\sum_{k}\frac{\tanh(\beta{\bar \varepsilon}_{k}/2)}
{2{\bar \varepsilon}_{k}}, \nonumber\\
C_{1}=\frac{Ja^2}{V_{0}^2}+C,
  \end{eqnarray}
  and $C$  is given by Eq.(\ref{expnegU}).
  The complete expansion of the functional (\ref{pairhopfunc}) to the 
  4th order is fully specified  by 
  the above expressions
   together with the coefficient ${\bar d}$ given by
  Eq.({\ref{coeffd}) and $b$ given by Eq.({\ref{coeffb}).
  
\section{Gaussian Fluctuations and the T--Matrix Approach}
In this Section we will analyze the contribution from the Gaussian fluctuations
to the free energy and the self-consistent T-matrix approach.
The effect of  Gaussian fluctuations of the order
parameter is obtained by keeping the quadratic terms  only in the full
functional.
The corresponding contribution to the thermodynamic potential is simply given by
\begin{equation}\label{gauss}
\beta {\cal F}^{(2)}= \frac{1}{N}\sum_{{\bf q} n}
\ln[1+|U|\chi({\bf q},i\varepsilon_{n})],
\end{equation}
where 
\begin{equation}\label{freeparticle}
\chi({\bf q},i\varepsilon_{n})=\frac{-1}{\beta N}
\sum_{ {\bf k} m}
 G_{0}({\bf k},i\omega_{m})
G_{0}({\bf q-k},i\varepsilon_{n}-i\omega_{m}),
\end{equation}
with $G_{0}({\bf k},i\omega_{n})^{-1}=i\omega_{n}-(\varepsilon_{k}-\mu)$,  
and   is given by Eq.(\ref{free}).

The BCS pairing instability corresponds to a singularity for 
${\bf q}=0, \omega=0$ in  ${\cal F}^{(2)}$, and the condition (the Thouless criterion)
$1+|U|\chi(0,0)=0$ gives $T_{c}$.

Equations (\ref{gauss},\ref{freeparticle}) can be used  to 
 form  a
fully renormalized, conserving approximation for the free energy and Green
function \cite{luttinger,baym,serene}.
Let the functional $\Phi[G]$  be defined as
\begin{eqnarray}
\Phi[G]=\frac{1}{\beta N}\sum_{{\bf q} n}
\ln[1-|U|{\bar \chi}({\bf q},i\varepsilon_{n})],\\
{\bar \chi}({\bf q},i\varepsilon_{n})=
\frac{1}{\beta N}\sum_{{\bf k} m}
G({\bf k},i\omega_{m})
 G({\bf q-k},i\varepsilon_{n}-i\omega_{m}),\label{chi}
\end{eqnarray}
where $G({\bf k},i\omega_{n})$ is the full Green function and  we use 
conventional definition of the pairing susceptibility Eq.(\ref{chi}).

The corresponding self-consistent approximation for the free energy is given by
the functional
\begin{eqnarray}
{\cal F}[\Sigma,G]= - \frac{2}{\beta N}
\sum_{{\bf k} n}\exp(i\omega_{n}\eta)\left\{
\Sigma({\bf k},i\omega_{n})G({\bf k},i\omega_{n})+
\ln[-G_{0}({\bf k},i\omega_{n})^{-1} +\Sigma({\bf
k},i\omega_{n})]\right\}\\\nonumber 
+\Phi[G],
\end{eqnarray}
evaluated at its stationary point with respect to variations 
of $G$ and $\Sigma$. At this stationary point $G, \Sigma,$ and $\Phi$ 
are related
by:
\begin{eqnarray}\label{dyson}
G({\bf k},i\omega_{n})=
[G_{0}({\bf k},i\omega_{n})^{-1}-\Sigma({\bf k},i\omega_{n})]^{-1},\\ 
\Sigma({\bf k},i\omega_{n})=\frac{1}{2}\frac{\delta \Phi[G]}
{\delta G({\bf k},i\omega_{n})}. 
\end{eqnarray}

Evaluation of $\Sigma({\bf k},i\omega_{n})$ yields
\begin{eqnarray}\label{Tapprox}
\Sigma({\bf k},i\omega_{n})=
\frac{1}{\beta N}\sum_{{\bf q} m}T({\bf q},i\varepsilon_{m})
G({\bf q - k},i\varepsilon_{m}-i\omega_{n}),
\end{eqnarray}
where 
\begin{eqnarray}\label{Tmatrix}
T({\bf q},i\varepsilon_{n})=\frac{-|U|}
{1-|U|{\bar \chi}({\bf q},i\varepsilon_{n})}
\end{eqnarray}
is the T-matrix.

The electron concentration  as a function of $T$ and $\mu$ is obtained 
as $n(T,\mu)=-\partial {\cal F}/\partial\mu$. By using stationary properties of
${\cal F}$  one gets the expected result
\begin{equation}\label{density}
n(T,\mu)= \frac{2}{\beta N}\sum_{{\bf k} n}\exp(i\omega_{n}\eta)
G({\bf k},i\omega_{n}).
\end{equation}
The above equations 
(\ref{dyson},\ref{Tapprox},\ref{Tmatrix},\ref{chi},\ref{density}) constitute 
the system of  self-consistent T-matrix 
equations and they were recently analyzed numerically 
for the attractive Hubbard model in two dimensions (mostly in low
concentration regime) 
 \cite{zurich}.

\section{Self-consistent approach to the superconducting 
fluctuations}
 \subsection{Hartree theory and  variational principle approach} 
 We would like to derive the best Gaussian form for the 
 superconducting 
 fluctuations. 
 Let us seek for a trial functional $\Psi_{T}$ in the following form 
 \begin{equation}
 \Psi_{T}[\Delta_{q}]=\sum_{q}A_{q}|\Delta_{q}|^2,
 \end{equation}
 where $q=({\bf q},i \varepsilon_{n})$.
 Application of the Feynman variational principle allows us to write 
 the corresponding variational free energy $F_{0}$
  \begin{equation} \label{Feynman}
  \beta F_{0}=-\ln\int D^{2}\Delta e^{-\Psi_{T}[\Delta_{q}]} +
  \langle\Psi-\Psi_{T}\rangle_{T}~~,
  \end{equation} 
  where $\left\langle \cdots \right\rangle_{T}=
\int D ^{2}\Delta \exp(-\Psi_{T}[\Delta_{q}])
(\cdots)/\int D^{2}\Delta \exp(-\Psi_{T}[\Delta_{q}])$.  
To evaluate  the RHS of Eq.(\ref{Feynman}) 
  we consider two cases:
   \begin{enumerate}
   \item We use expansion of the exact functional $\Psi$ truncated at the 4th 
   order term.
   \item We use the exact form of $\Psi$ given by Eq.(\ref{exact})
   \end{enumerate} 
   In the first case we  use $\Psi$ as given by Eq.(\ref{exp})
    and after exploring the properties of  Gaussian integrals we obtain  for $F_{0}$
 \begin{eqnarray}
 \beta F_{0}=-\ln \int D^{2}\Delta
 e^{-\Psi_{T}[\Delta_{q}]}+
 \sum_{q}\left[\frac{1}{|U|}+\chi(q)-A_{q}\right]\langle|\Delta_{q}|^2\rangle\nonumber\\+
 \frac{1}{2N\beta}\sum_{p_{1},p_{2}}
 \Gamma^{+-+-}(p_{1},p_{2},p_{2})\langle|\Delta_{p_{1}}|^2\rangle\langle|\Delta_{p_{2}}|^2\rangle\nonumber\\+
 \frac{1}{2N\beta}\sum_{p_{1},p_{2}}\Gamma^{+-+-}(p_{1},p_{2},p_{1})\langle|\Delta_{p_{1}}|^2\rangle\langle|\Delta_{p_{2}}|^2\rangle~~,
 \end{eqnarray} 
 \noindent
 where $\langle|\Delta_{p}|^2\rangle$ is the functional average with $\Psi_{T}$.  
 The best coefficients $A_{q}$ are obtained from the stationary  
 condition $\frac{\delta(\beta F_{0})}{\delta A_{q}}=0$.
  With the  notation 
  $t_{q}=<|\Delta_{q}|^2>$ for the mean square amplitude of the 
  order parameter (fluctuating in space and time), 
  we obtain 
  \begin{equation}
  A_{q}=\frac{1}{|U|} +\chi(q)+\frac{1}{2N\beta}
  \sum_{p}\left[\Gamma^{+-+-}(q,p,p)+
  \Gamma^{+-+-}(p,q,p)\right]t_{p}.
  \end{equation}
   Since $\Psi_{T}$ is the Gaussian functional 
   then $t_{q}=\frac{1}{A_{q}}$, and 
   we obtain the following self-consistent equation for $t_{q}$: 
   \begin{equation}\label{Hartree}
   t_{q}=\left[\frac{1}{|U|}+\chi(q)+\frac{1}{2N\beta}\sum_{p}\left[\Gamma^{+-+-}(q,p,p)+\Gamma^{+-+-}(p,q,p)\right]t_{p}\right]^{-1}~~.
   \end{equation} 
   If we neglect  the frequency dependence and perform small momentum expansion for $\chi(q)$ and also neglect 
  the  momentum dependence of the $\Gamma$ terms in (\ref{Hartree}), 
   then  we  recover the Hartree equation for the amplitude fluctuations used in Section VI.
   \subsection{Fully self-consistent approach}
   In the second case we use the complete functional (\ref{exact}) 
   to evaluate the trial free energy (\ref{Feynman})\cite{everts}. 
   By making the substitution $\Delta_{q}=A^{-1/2}_{q}\xi_{q}$, one 
   gets
   \begin{equation}
   \beta F_{0}=-\sum_{q}(1-\ln(|U| A_{q}))+\int D^2\xi_{q}e^{-\sum_{q}|\xi_{q}|^2}\Psi[A_{q}^{-1/2}\xi_{q}].
   \end{equation}
   The stationary condition $\frac{\delta(\beta F_{0})}{\delta A_{q}}=0$, after 
integrating by parts, yields the following  expression for
$A_{q}$
\begin{equation}
\label{full}
A_{q}=
\left\langle\frac{\delta^2\Psi}{\delta\Delta_{q}
\delta\Delta^{*}_{q}}\right\rangle_{\rm {T}} ~,
 \end{equation}
i.e. the optimal quadratic functional coefficients 
are self-consistently averaged 
second derivatives of the true functional.
We note that the random phase approximation (RPA) result is obtained if the second derivative is 
evaluated at  the origin:
\begin{equation}
A_{q}=
\frac{\delta^2\Psi}{\delta\Delta_{q}\delta\Delta^{*}_{q}}|_{\Delta=0},
\end{equation}
     and it simply yields Eq.(25) with  
     the free particle--particle susceptibility (\ref{free}). 
     By using Eq.(\ref{full}) and the free energy functional Eq.(\ref{exact}),
     after tedious  manipulations (similar to those  in Ref.\cite{Hertz}), 
     one arrives at
     \begin{eqnarray}
     \frac{{\delta}^2}{\delta\Delta_{q}\delta\Delta^{*}_{q}}\left(-Tr\ln(1-\hat{\Delta}\hat{\cal G}_{0})\right)~=
     ~\sum_{k,k'}{\cal{G}}^{11}_{k,k'}\{\Delta\}{\cal{G}}^{22}_{k+q,k'+q}\{\Delta\},
     \end{eqnarray}
where  ${\cal{G}}^{\alpha\alpha}_{kk'}\{\Delta\}$, $\alpha=1,2$, 
denotes the diagonal matrix element of the space-time Fourier transform of  
$\hat{\cal{G}}_{i,j}(\tau,\tau',\{\Delta\})$ given by Eq.(\ref{local}). 
We note that $\hat{{\cal{G}}}\{\Delta\}=
\hat{{\cal{G}}}_{0}(1-\hat{\Delta}\hat{\cal{G}}_{0})^{-1}$ 
is the  electron propagator in the random potential $\Delta_{i}(\tau)$. 
The expression for the coefficient 
$A_{q}=A({{\bf q}},i\epsilon_{n})$ is then given by
     \begin{equation}
     \label{gaussprop}
     A({\bf q},i\epsilon_{n})=\frac{1}{|U|}+\bar{\chi}({\bf q},i\epsilon_{n}),
     \end{equation}
     where
     \begin{eqnarray}
     \bar{\chi}({\bf q},i\epsilon_{n})=
     \frac{1}{\beta N}\sum_{{\bf k},{\bf k'},m}}
     \langle{{\cal{G}}^{11}
     ({\bf {k}}, {\bf {k'}},i\omega_{m};\{\Delta\}){\cal{G}}^{22}
     ({\bf {k+q}}, {\bf {k'+q}},i\omega_{m+n};\{\Delta\})
     \rangle_{\rm {T}}
     \end{eqnarray}
     Therefore, we have derived a self-consistent theory  of superconducting 
     fluctuations in which  the paring susceptibility 
     $\bar{\chi}({\bf k},(i\epsilon_{n})$ 
     is given by the averaged particle--particle propagator in the presence 
     of the random fields.  
     In this approach the particle--particle propagator in  
     Eq.(\ref{gaussprop}) 
     is just the product of two local Green 
     functions self-consistently averaged 
     with the Gaussian weight $\exp(-A_{q}|\Delta_{q}|^2)$. 
     (To lowest order in $|U|$, when we 
     ignore the random fields, $\bar{\chi}$ reduces to the 
     bare $\chi$, and we recover the RPA  result).
     The above theory is very similar 
     in the spirit to the paramagnetic polaron theory 
     in itinerant-electron magnetism\cite{Hertz} 
     and can be used to describe incoherent local pair  
     state. Finally, we note that the presented theory 
     when resorting to the static approximation 
     yields a theoretical foundation for the approach in which the electronic DOS is 
     evaluated with the assumption of  existence the local order parameter
      amplitude above $T_{c}$ and averaged with the Gaussian weight.
\section{Amplitude fluctuations above 
      $T_{c}$ in the Ginzburg-Landau theory 
      and incoherent local pair state}  It is of interest to study further the   
       fluctuation effects (and pairing correlations)  and their role in establishing pair formation 
       above T$_{c}$, within the 
       conventional  Ginzburg-Landau theory. 
       From the previous analysis, one comes to the conclusion 
       that above T$_{c}$ there will be a region of parameters in 
       which there are pairs with amplitude and phase fluctuating 
       in space and time. Such a state, however, has no 
       long-range phase coherence which is  necessary to establish the 
       superconducting state. This scenario is somewhat similar to a 
       local-moment 
       itinerant electron magnetism in transition metals and alloys, 
       in which the magnetic moments exist above transition 
       temperature but their direction is random.  
       A possibility of such a state in 
       the negative $-U$ Hubbard model has been considered 
       earlier\cite{micnas90} 
       to describe the intermediate region 
       between BCS and the local pair regime. 
       Such an incoherent local pair state has also been analyzed by 
       Gyorffy et al \cite{gyorffy,park} within the coherent potential 
       approximation to the 
       negative $-U$ Hubbard model with the use of a constant density of states. 
    
       In order to show the existence of a finite gap amplitude above 
       $T_{c}$,  we consider 
        the standard Ginzburg-Landau functional (G-L) and apply 
        the 
        Hartree approximation to treat the interaction between 
        superconducting fluctuations\cite{park}.\\
        In zero magnetic field one has
        \begin{equation}
        F[\Psi({\bf r})]=\int d{\bf r}\left [ \bar{a(t)}|\Psi({\bf r})|^{2}+\bar{b}|\Psi({\bf r})|^{4} +
        \frac{\hbar^{2}}{4m}|\nabla\Psi({\bf r})|^2\right ],
\end{equation} 
\noindent
where $t=T/T_{c}$ and $m$ is an effective electron mass.
The superconducting order parameter $\Psi({\bf r})$ can be related to 
the gap parameter 
$\Delta({\bf r})$, for example, via 
the Gorkov relation:
$\Psi({\bf r})=(\frac{7\zeta(3)n}{8(\pi k_{B}T_{c})^{2}})^{1/2}\Delta({\bf r})$, with $n$ the 
number of electrons per unit volume. With the Fourier transform to reciprocal
 space we have 
\begin{equation}
F[\Delta_{{\bf k}}]/\Omega  = 
\sum_{\bf k} \left[ [a(t)+ck^{2}]|\Delta_{\bf k}|^{2}+
b\sum_{{\bf p,q}}\Delta_{\bf k}\Delta_{\bf p}^{\star}\Delta_{\bf q}\Delta _{\bf k-p+q}^{\star}\right],
\end{equation} 
\noindent
where $\Omega$ is the volume of the system.
Assuming that $b$ and $c$ are temperature independent and performing 
the simplest 
quadratic factorization of the last  
term we obtain the gaussian form of the G-L functional: 
\begin{equation}
F[\Delta_{{\bf k}}]/\Omega  = \sum_{\bf k} \left[ a(t)+ ck^{2}+
2b \sum_{{\bf q}}<|\Delta_{q}|^2>\right]|\Delta_{\bf k}|^{2}.
\end{equation} 
In the above equation the expectation value 
has to be calculated as the functional average: 
\begin{equation}
\langle....\rangle =\frac{\int D^2 \Delta (...) \exp[-F[\Delta_{\bf k}]/(k_{B}T)]}{\int D^2 \Delta \exp[-F[\Delta_{\bf k}]/(k_{B}T)]}.
\end{equation} 
The  mean square gap amplitude $
\Delta^{2}= \sum_{{\bf k}}<|\Delta_{{\bf k}}|^{2}>$ is then 
given by the following self--consistent equation 
\begin{equation}
\Delta^2=\frac{k_{B}T}{\Omega}\sum_{k}\left[\frac{1}{a(t)+ ck^{2}+2b\Delta^{2}}\right].
\end{equation} 
In 2D we obtain 
\begin{equation}
\ln(1+w/x)=({\tilde d}/t)(x+1-t),
\end{equation} 
\noindent
where the following relations have been used 
\begin{eqnarray*}
a(t)=a(0)(1-t),~ t=T/T_{c},~ |\Delta_{0}|^2=-a(0)/2b, \\x=\Delta^2/|\Delta_{0}|^2 -1+t, 
 w=(k_{c}\xi_{0})^2, \\{\tilde d}=\frac{4\pi|a(0)|\xi_{0}^2|\Delta_{0}|^2}{k_{B}T_{c}},~
  a(0)=-\frac{7\zeta(3)\hbar^2 n|\Delta_{0}|^2}{8\pi m(k_{B}T_{c})^2}.
\end{eqnarray*} 
In 3D case we get 
\begin{equation}w^{\star}-\sqrt{x}\arctan(\frac{w^{\star}}{\sqrt{x}})=(d^{\star}/t)(x+1-t),
\end{equation} 
where 
\begin{eqnarray*}w^{\star}=k_{c}\xi_{0},~~d^{\star}=
\frac{7\zeta(3)\hbar^2 n|\Delta_{0}|^2\xi_{0}}{16m(k_{B}T_{c})^3}.
\end{eqnarray*} 
\noindent$\xi_{0}$ and $\Delta_{0}$ 
are the zero temperature coherence length 
and the zero temperature energy gap, respectively, 
$k_{c}$ denotes the momentum 
cutoff. 

We have solved these equations with the 
use of parameters appropriate for  YBa$_{2}$Cu$_{3}$O$_{7}$, 
taking $|\Delta_{0}|=20 meV$, $T_c$~=~92 K, $n~=~6*10^{14}cm^{-2}$ (for 2D case)
 and $n~\sim 10^{21}cm^{-3}$ (for 3D case), $m~=~5m_{e}$, $k_{c}~=~\xi_{0}^{-1}$. 
The result is illustrated in Figs.1-2 for several values of ${\tilde d}$ and $d^{\star}$. 
For both 2D and 3D cases for T above $T_{c}$, $\Delta$ 
smoothly decreases and saturates at a finite value. Depending on the values of 
${\tilde d}$ and $d^{\star}$, 
$\Delta$ is of order$(0.1-0.2)\Delta_{0}$ for 2D case and of order $(0.05-0.125)\Delta_{0}$ in 3D.\\
The parameter $k_{c}$ is not precisely defined in the Ginzburg-Landau 
theory and we consider it to vary. If we allow  randomness of $\Delta({\bf r})$ 
down to the atomic 
scale, then  $k_{c}$ can be taken as an inverse of the  lattice constant 
$(4\AA)$,
giving w = 9 (2D) and $w^{\star}=3$ (3D). In this case, 
we observe (Figs.1-2) that after an initial decrease, $\Delta/\Delta_{0}$  starts to 
increase and saturates at  larger values. 
The gap magnitude is of the order $(0.15-0.275)\Delta_{0}$ in 2D and $(0.1-0.2)\Delta_{0}$ in 3D,
depending on the values of the  parameters ${\tilde d}$ and $d^{\star}$. 
Thus if we lower $k_{c}$, 
increasing temperature is  more effective in making larger thermal  fluctuations in the 
amplitude of the local
 order parameter $\Delta({\bf r})$. 
This characteristic behavior  is very similar to that known 
in the theory of local moment magnetism\cite{moriya}.    
 The above calculations carried out  at the phenomenological Ginzburg-Landau 
 level
 clearly illustrate  the importance of the fluctuations above $T_c$ and point out 
 toward a  possibility of existence of a state with 
 incoherent pairs. In the incoherent local pair  state the one-particle 
 density of states will develop a pseudogap or real gap 
 depending on the electron density and  
 the strength of the  attractive interaction.  
\acknowledgements   
A part of this work has been initiated  at 
IBM Research Division--R\"uschlikon, Switzerland. Thanks are due to H. Beck, S.
Robaszkiewicz  and
 T. Schneider for helpful discussions. 

\newpage

\begin{figure}[htp]
\centerline{
\psfig{figure=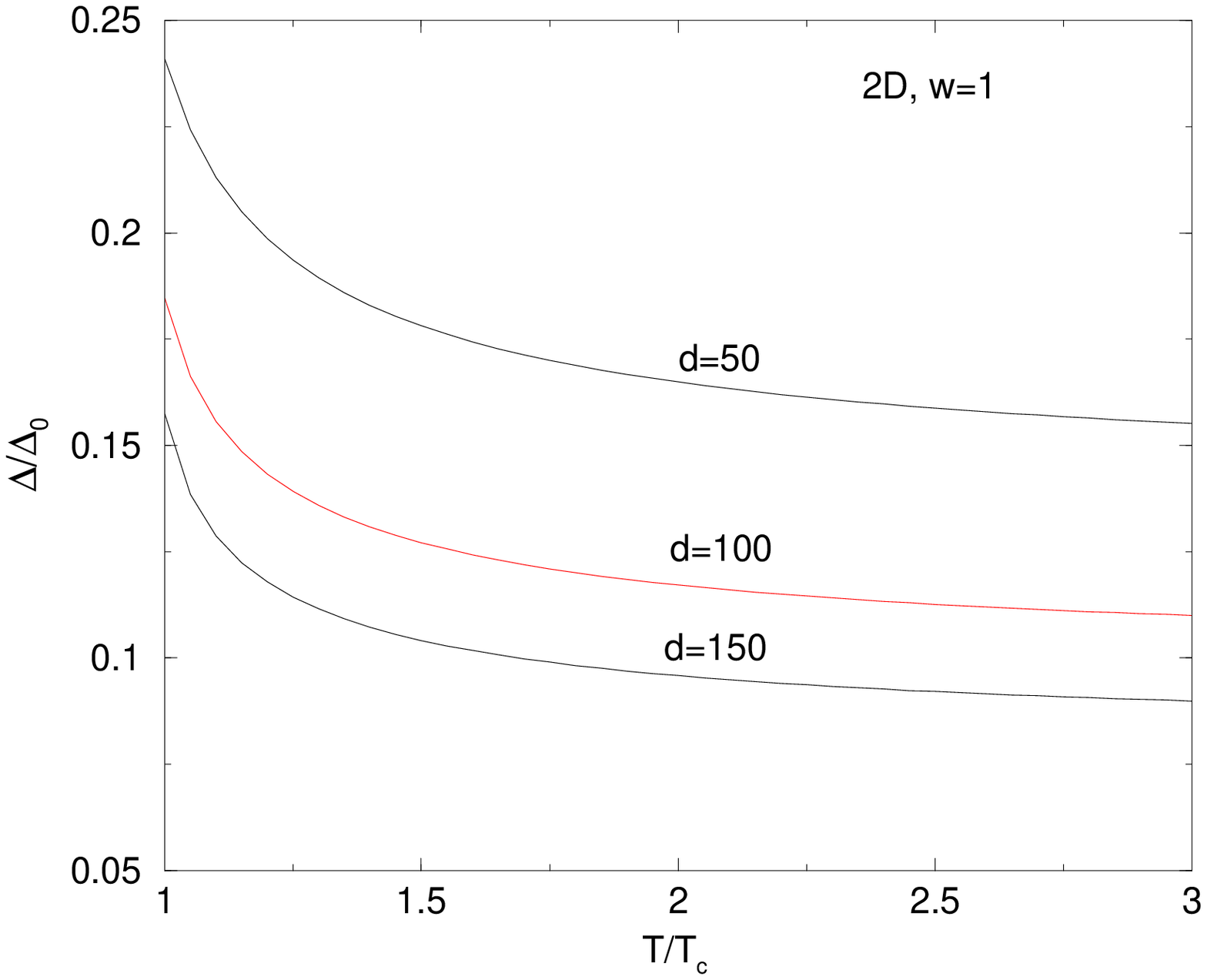,width=8 cm,height=8cm}
\psfig{figure=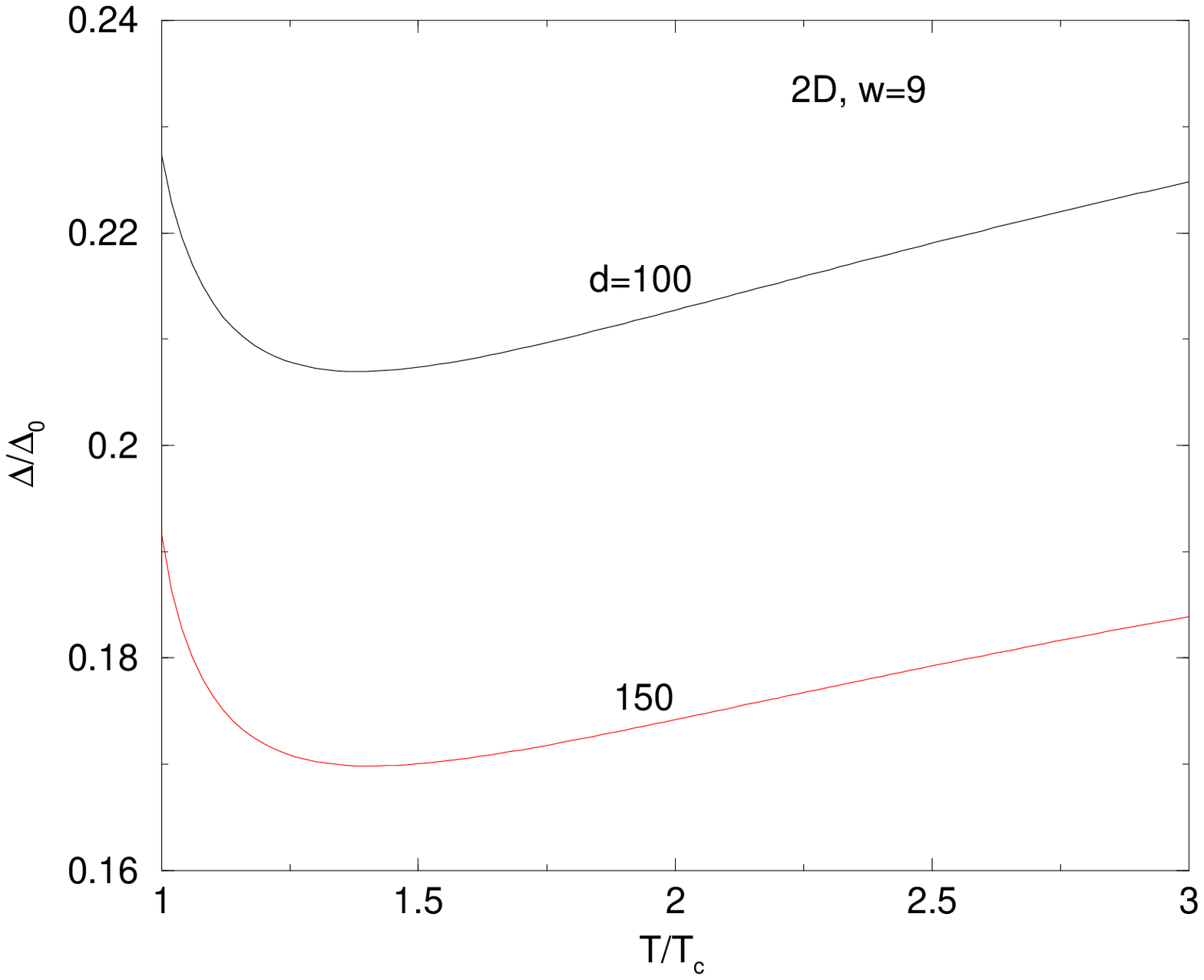,width=8cm,height=8cm}
}
\caption{$\frac{\Delta}{\Delta_{0}}$ versus $T/T_{c}$  in two dimensions for
two different values of the cutoff $k_{c}$.}

\end{figure}

\begin{figure}[htp]
\centerline{
\psfig{figure=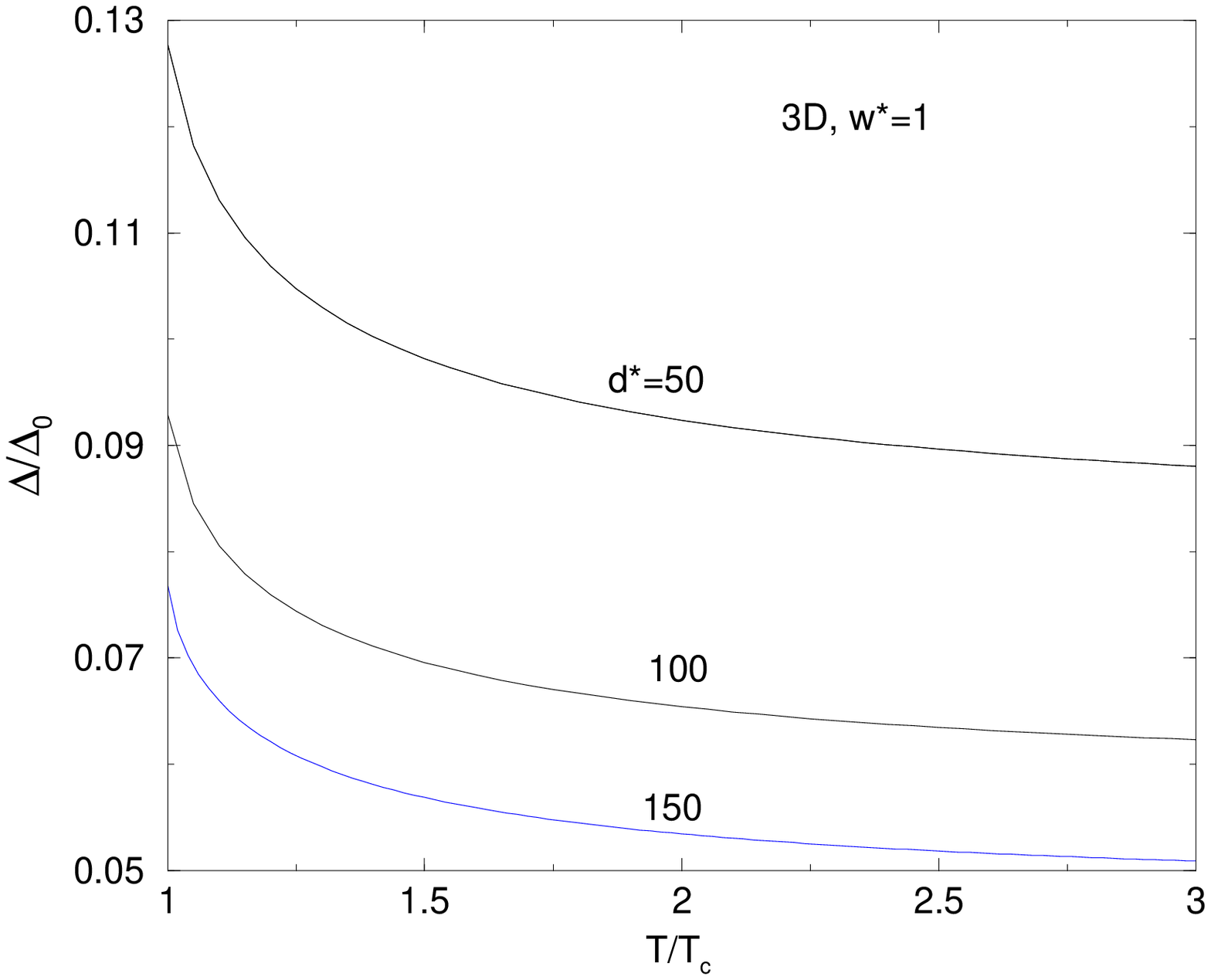,width=8cm,height=8cm}
\psfig{figure=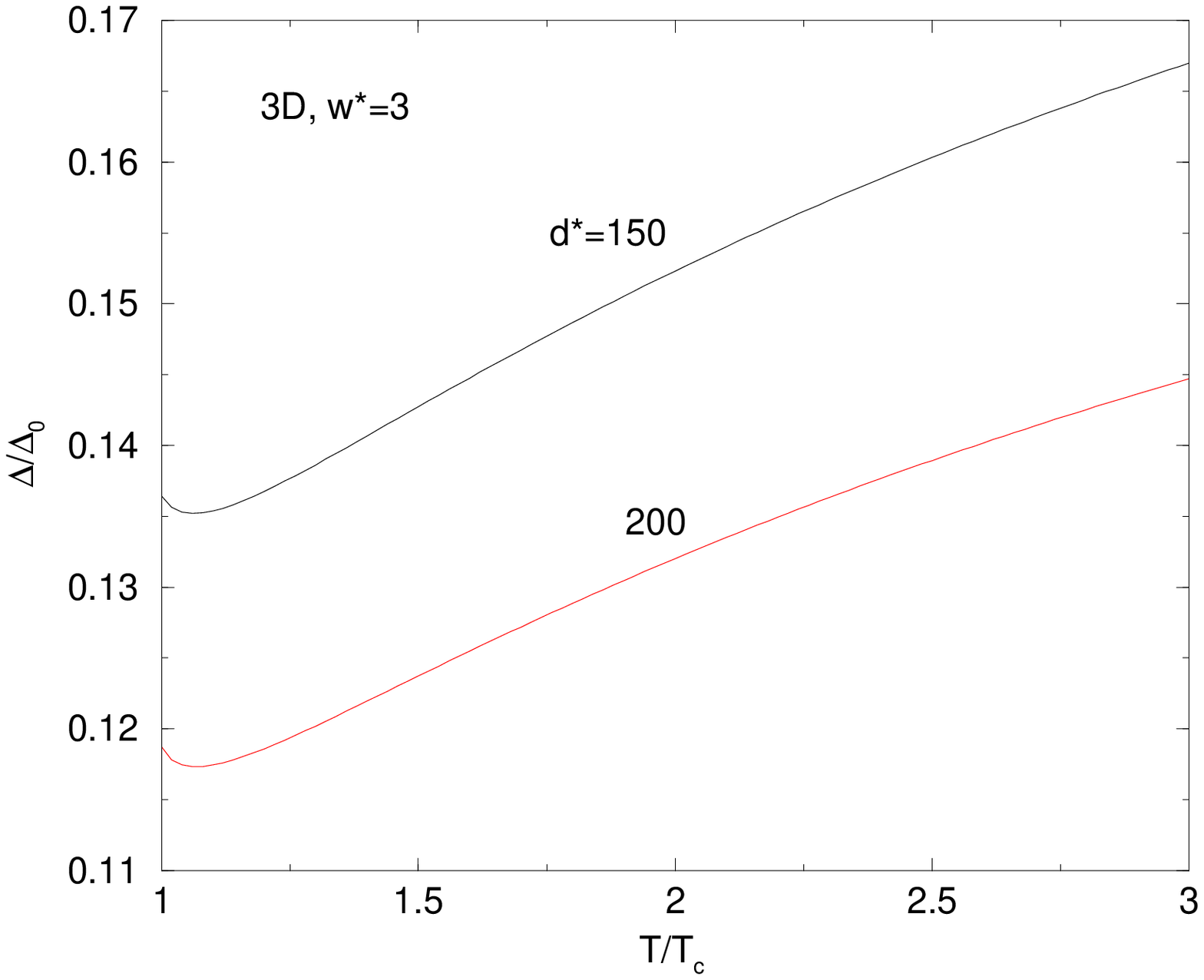,width=8cm,height=8cm}
}
\caption{$\frac{\Delta}{\Delta_{0}}$ versus $T/T_{c}$  in three dimensions for
two different values of the cutoff $k_{c}$.}

\end{figure}

\begin{references}
\bibitem{micnas90} R. Micnas, J. Ranninger and S. Robaszkiewicz, Rev.\ Mod.\ Phys. {\bf 62}, 113 (1990).
\bibitem{uemura91} Y.J. Uemura, L. P. Le, G. M. Luke, B. J. Sternlieb, W. D. Wu, J. H. Brewer, 
T. M. Riseman, C. L. Seaman, M. B. Maple, M. Ishikawa, D. G. Hinks, 
J. D. Jorgensen, G. Saito, and H. Yamochi, Phys. Rev. Lett. {\bf 66}, 2665 (1991);\\ 
Y. J. Uemura, A. Keren, L.P.Le , G.M. Luke, B.J. Sternlieb, W.D Wu, 
J.H.Brewer, R.L. Whetten , S.M. Huang , S. Lin, R.B. Kaner, 
F. Diederich, S. Donovan, G. Gruner, K. Holczer , Nature {\bf 352}, 605 (1991).
\bibitem{schneider92} T. Schneider and H. Keller, Phys.\ Rev.\ Lett.\ {\bf 69}, 3374 (1992); 
T. Schneider and H. Keller, Int. J. Mod. Phys.{\bf B 8}, 487 (1994).
\bibitem{NATO}R. Micnas and S. Robaszkiewicz, in 
{\it "High-$T_{c}$
Superconductivity 1996: Ten Years after the Discovery"}
(E. Kaldis, E. Liaropkapis and K.A. M\"uller, eds.)  NATO ASI E {\bf 343}, p.31 (1997)
(Kluwer) and Refs. therein.
\bibitem{orenstein} J. Orenstein and A.J. Millis, Science {\bf 288}, 468 (2000).
\bibitem{eagles} D. M. Eagles, Phys. Rev. {\bf 186}, 456 (1969).
\bibitem{leggett} A. J. Leggett, in "Modern Trends in the Theory of Condensed Matter", 
edited by A. P\c{e}kalski and J. Przystawa (Springer Verlag, Berlin 1980),p.13..
\bibitem{rob81} S. Robaszkiewicz, R. Micnas, and K.A. Chao, Phys.\ Rev. {\bf B 23}, 1447 (1981); ibid.{\bf B 24}, 1579 (1981); {\bf B 24}, 4018 (1981); {\bf B 26}, 3915 (1982).
\bibitem{noz85} P. Nozieres and S. Schmitt-Rink, J.Low Temp.\ Phys.,{\bf 59}, 195 (1985).
\bibitem{miyake} K. Miyake, Prog. Theor. Phys. {\bf 69}, 195 (1983).
\bibitem{ran} M. Randeria, J.-M. Duan and L.-Y. Shieh, Phys.\ Rev.\ Lett. {\bf 62}, 981 (1989); Phys. \ Rev. {\bf B 41}, 327 (1990).
\bibitem{melo}C.A.R. Sa de Melo, M. Randeria and J.R. Engelbrecht,
Phys. Rev. Lett. {\bf 71}, 3202 (1993).
\bibitem{Haussmann}R. Haussmann, Phys. Rev. {\bf B 49}, 12975(1994).
\bibitem{chicago}J. Maly, B. Janko and K. Levin, Phys . Rev. {\bf B 59},
1354(1999); Q. Chen, I. Kosztin, B. Janko and K. Levin, Phys. Rev. {\bf
B59}, 7083 (1999); I. Kosztin, Q. Chen, Y. Kao and K. Levin, Phys. Rev. {\bf
B61}, 11 662 (2000).
\bibitem{Varma et al} S. Schmitt-Rink, C.M. Varma and A. E. Ruckenstein, 
Phys. Rev. Lett. {\bf 63}, 445 (1989).
\bibitem{Traven} S.V. Traven, Phys. Rev. Lett. {\bf 73}, 3451 (1994). 
\bibitem{randeria94} M. Randeria, in {\it "Bose-Einstein Condensation"} edited by A. Griffin, D. 
Snoke and S. Stringari (Cambridge University Press, 1995), p.355.
\bibitem{TKRM} R. Micnas and T. Kostyrko, in
{\it "Recent Progress in High Temperature Superconductivity"},
(J. Klamut, B.W. Veal, B.M. Dabrowski, P.W. Klamut and M. Kazimierski, eds) Lecture Notes in
Physics vol. {\bf 475}, 221 (1996)(Springer Verlag, Berlin, Heidelberg) and Refs. therein.
\bibitem{Hubbard} J. Hubbard, Phys. Rev. Lett. {\bf 3}, 77 (1959);\\
R.L. Stratonovich, Sov. Phys. Doklady {\bf 2}, 371 (1959).
\bibitem{zwerger}M. Drechsler and W. Zwerger, 
Ann. Phys.(Leipzig) {\bf 1},15 (1992).
\bibitem{pieri} For a recent  analysis 
of the mapping of  fermionic system with attraction  
to the composite boson limit  see   P. Pieri and G.C. Strinati,
Phys. Rev. {\bf B 61}, 15 370(2000);
 F. Pistolesi and G.C. Strinati, ibid. {\bf B 53}, 15168(1996);
T. K. Kope\'c and H. Umezawa, ibid. {\bf B 47}, 8923 (1993).
\bibitem{Yamada}A. Tokumitu, K. Miyake and K. Yamada,  Phys. Rev. {\bf B 47},
11988 (1993). 
\bibitem{micnas92}R. Micnas and S. Robaszkiewicz, Phys. Rev. 
{\bf B 45}, 9900 (1992).
\bibitem{ukrainians}E.V. Gorbar, V.M. Loktev and S.G. Sharapov,
ITP-95-6E preprint of the N.N. Bogolyubov Institute for Theoretical Physica,
Kiev,  Ukraine, Physica {\bf C 257},355 (1996).
\bibitem{rob} W. Czart and S. Robaszkiewicz,Phys. Rev. {\bf B 64}, 104511
(2001) and Refs. therein.
\bibitem{luttinger} J. M. Luttinger and J.C. Ward, Phys. Rev. {\bf 118}, 1417 
(1960).
\bibitem{baym} G. Baym, Phys. Rev. {\bf 127}, 1391 (1962).
\bibitem{serene} J.W. Serene, Phys. Rev. {\bf B40}, 10873(1989).
\bibitem{zurich} R. Micnas, M. H. Pedersen, S. Schafroth, T. 
 Schneider, J.J. Rodriguez-Nunez and H. Beck, 
 {\it Phys.Rev.} {\bf B 52}, 16223 (1995); M. H. Pedersen, J.J. Rodriguez-Nunez,
H. Beck, T. Schneider and  S. Schafroth, Z. Physik, {\bf B 103}, 21 (1997);  see also J. J. Deisz, D. W. Hess and 
 J.W. Serene, Phys. Rev. Lett. {\bf 80}, 373 (1998);
 W. Keller, W. Metzner and U. Schollwock, Phys. Rev. {\bf B 60},3499(1999);
 D. Rohe and W. Metzner, Phys. Rev. {\bf B 63}, 224509(2001); B. Kyung, Phys. Rev. {\bf B 63}, 014502 (2001).
\bibitem{Hertz} J.A. Hertz and M. Klenin, Phys.Rev. {\bf B 10}, 1084 (1974).
\bibitem{everts} For an early study of variational treatment see H.-U. Everts,
Z. Physik {\bf 199}, 211 (1967).
\bibitem{gyorffy} B.L. Gyorffy, J.B. Staunton and G.M. Stocks, 
Phys. Rev. {\bf B 44}, 5190 (1991).
\bibitem{park} K.A. Park and R. Joynt, Phys. Rev. {\bf B 48},16833 (1993).
\bibitem{moriya} T. Moriya, 
{\it {Spin Fluctuations in Itinerant Electron Magnetism}} (Springer Verlag,\\ 
Berlin,1985), p.153.
\end{references}
\end{document}